\documentclass[aps,pra,showpacs,floatfix,superscriptaddress, twocolumn, showkeys]{revtex4-1}

\usepackage{amsmath,amsthm,verbatim,amssymb,amsfonts,amscd, graphicx}
\usepackage{dsfont}
\usepackage{graphics}
\usepackage{xcolor}
\usepackage{soul}
\usepackage{mathtools}
\usepackage{mathdesign}
\usepackage{braket}
\usepackage[utf8]{inputenc}
\usepackage[english]{babel}
\usepackage[T1]{fontenc}
\usepackage{babelbib}
\usepackage{url}
\usepackage[bottom]{footmisc}

\def\<{\langle}
\def\>{\rangle}

\def\set#1{{\sf #1}}

\def\map#1{{\mathcal{#1}}}

\def\Tr{\operatorname{Tr}}

\theoremstyle{plain}

\theoremstyle{definition}

\begin{document}

\title{Deterministic Twirling with Low Resources}
\author{David Jakob Stonner}
\address{Institut f\"ur Angewandte Physik, Technische Universit\"at Darmstadt,Hochschulstra{\ss}e 4a, D-64289 Darmstadt, Germany}
\author{Jaroslav Kysela}
\thanks{Present address: Institute for Quantum Optics and Quantum Information (IQOQI), Austrian Academy of Sciences, Boltzmanngasse 3, 1090 Vienna, Austria}
\address{Department of Physics, Faculty of Nuclear Sciences and Physical Engineering, Czech Technical University in Prague, B\v{r}ehov\'a 7, 115 19 Praha 1-Star\'e M\v{e}sto, Czech Republic}
\author{Graeme Weir}
\email{graeme.weir@fjfi.cvut.cz}
\address{Department of Physics, Faculty of Nuclear Sciences and Physical Engineering, Czech Technical University in Prague, B\v{r}ehov\'a 7, 115 19 Praha 1-Star\'e M\v{e}sto, Czech Republic}
\author{Jaroslav Novotn\'y}
\address{Department of Physics, Faculty of Nuclear Sciences and Physical Engineering, Czech Technical University in Prague, B\v{r}ehov\'a 7, 115 19 Praha 1-Star\'e M\v{e}sto, Czech Republic}
\author{Gernot Alber}
\address{Institut f\"ur Angewandte Physik, Technische Universit\"at Darmstadt,Hochschulstra{\ss}e 4a, D-64289 Darmstadt, Germany}
\author{Igor Jex}
\address{Department of Physics, Faculty of Nuclear Sciences and Physical Engineering, Czech Technical University in Prague, B\v{r}ehov\'a 7, 115 19 Praha 1-Star\'e M\v{e}sto, Czech Republic}

\begin{abstract}
Twirling operations, which average a quantum state with respect to a unitary subgroup, have become a frequently-employed tool in quantum information processing. We investigate the efficient implementation of twirling operations with minimal resources, without necessitating the ability to perform all possible unitary operations on the quantum system of interest. We present a general algebraic method allowing us to choose a set of - typically very few - unitary operators which, when applied randomly and repeatedly, produce the given twirling operation exponentially quickly. The method is applied to twirling operations for bipartite quantum systems with respect to the unitary group $\map U(d)\otimes\map U(d)$, an essential ingredient in entanglement distillation protocols. In particular, we provide a complete classification of sets of unitary operators capable of performing twirling on two qubits. Moreover, we construct a generic set containing at most three unitary operators achieving the twirling operation for a general two-qudit system.
\end{abstract}


\maketitle
\section{Introduction}

Entanglement, due to its delicate quantum correlations,
lies at the heart of numerous quantum information protocols \cite{BennettPurification1996,Delgado,Werner2001,
Lee2003,Hayashi2008}.
Therefore, our ability to prepare, distribute, and manipulate entanglement efficiently
is one of the key factors enabling a significant further advancement of quantum technology.
Unfortunately, entanglement is rather fragile and is affected easily even by small interactions
of quantum systems with their surroundings \cite{Novotny2011}. In order to counteract such destructive influences, powerful
entanglement distillation protocols have been designed allowing
remote participants to prepare, by local means, high-fidelity entangled states even from poorly-entangled shared
sources \cite{BennettPurification1996}. These protocols rely on the existence of efficient twirling
operations which are capable of converting multipartite quantum states
into so-called Werner states \cite{Werner}. In general, a twirling operation $\map T$ performs a group averaging over a subgroup $\cal G$ of the unitary group ${\cal U}(d)$, i.e.
\begin{equation}
\label{eq:def_general_twirling}
\map T(\rho)=\int_{\cal G} U \rho\,U^{\dagger} \mathrm{d}U,
\end{equation}
with $\rho$ and $\mathrm{d}U$ denoting an arbitrary quantum state and the Haar measure on $\cal G$, respectively. Furthermore, $d$ is the dimension of the underlying Hilbert space. For finite groups the integral simplifies to a sum over all group elements, normalised by the group order.
Twirling operations have interesting applications in numerous areas of quantum information processing, such as quantum distillation processes \cite{BennettPurification1996,Delgado},
theory of entanglement measures \cite{Werner2001,Lee2003,Hayashi2008},
resource theory of asymmetry \cite{Marvian2013,Marvian2016},
quantum reference frames \cite{Vaccaro2008,Skotiniotis2012},
quantum secret-sharing \cite{Gheorghiu2012},
data hiding \cite{DiVincenzo2002},
depolarization of quantum channels \cite{Dur2005}, and the
addressability of quantum gates \cite{Gambetta2012}.

In the standard approach, a general twirling operation is obtained by applying a sequence of $M$ randomly chosen unitary operators $\left(U_i\right)_{i=1}^{M}$ from the group $\cal G$ randomly to $M$ systems prepared in the same initial state $\rho$: i.e., the output state is given by the averaging operation $\map T_M$
\begin{equation}
\label{eq:def_random_unitary}
\rho' = \map T_M(\rho) = \frac{1}{M} \sum_{i=1}^{M} U_i \rho\,U_i^{\dagger}.
\end{equation}
Practical implementations of this straightforward procedure face two serious issues. Firstly, experimentally it must be possible to realise all required unitary operators from such a group reliably. In general this is a highly demanding task.
Secondly, even if this can be accomplished, typically
the convergence of such a straightforward procedure scales polynomially
with the number of steps (unitaries) involved \cite{Toth2007}. However, it was numerically demonstrated \cite{Toth2007} that with a clever choice of a few unitary
operators, not even necessarily from the group $\cal G$, which are applied randomly and repeatedly,
one can achieve an exponential speedup. Thus, instead of increasing the number of unitary operators, it is significantly more efficient to repeat an averaging operation $\map T_m$ with $m$ fixed but well-chosen unitary operators so that after $N$ iterations the quantum system is in the state 
\begin{equation}
\label{eq:def_iterative_approach}
\rho'= \left(\map T_m\right)^N (\rho).
\end{equation}
Such an iterative procedure can result in an exponential convergence towards the desired group averaging operation. Simultaneously it requires significantly less resources.
Motivated by this advantage of the latter procedure, in this paper we present a general algebraic method
addressing the crucial practical question of how to choose the set of unitary operators
$(U_i)_{i=1}^{m}$ whose repeated random application approximates a given group-averaging operation
exponentially quickly. In the following, this method will be applied not only to the twirling of two qubits by providing a characterisation of an arbitrary number of unitary operators achieving this goal, but we will also present generic sets containing at most three unitary operators which achieve twirling exponentially quickly in arbitrary finite-dimensional bipartite quantum systems.

\section{A general algebraic method} \label{sec:RUO}
We present a general method for determining the set of unitary operators
$(U_i)_{i=1}^{m}$ of a random unitary operation (RUO)
\begin{equation}
\label{eq:def_RUOs}
\map R(\rho) = \sum_{i=1}^{m} p_i U_i \rho\, U_i^{\dagger}    
\end{equation}
which efficiently approximates a given twirling operation 
\eqref{eq:def_general_twirling} by the iterative evolution
\eqref{eq:def_iterative_approach}. Thereby, the quantities
$p_i > 0$ with $i=1,...,m$ and $\sum_{i=1}^m p_i =1$ describe the probabilities with which the corresponding
unitary operator $U_i$ is applied.
The asymptotic regime of the generated iterative evolution is well-understood in terms of attractors of an attractor space $\set{Attr}\left(\map R\right)$ associated with the 
given RUO \cite{Novotny2009,NovotnyRUOs}. This attractor space is defined as the span of all eigenvectors of the map $\map R$ associated with eigenvalues of the asymptotic spectrum: i.e.,
\begin{equation}
\label{eq:attractor_space}
    \set{Attr}\left(\map R\right) \coloneqq \bigoplus_{\lambda \in \sigma_\mathrm{as}} \set{Ker}\left(\map R - \lambda \mathds{1} \right),
\end{equation}
with
the asymptotic spectrum $\sigma_\mathrm{as}$ being defined as the set of eigenvalues of $\map R$ with unit modulus. Once an orthonormal basis $X_{\lambda,i}$ of the individual attractor eigenspaces $\set{Ker}\left(\map R - \lambda \mathds{1} \right)$ with respect to the Hilbert-Schmidt scalar product $ (X,Y)_\mathrm{HS} = \Tr\{X^{\dagger}Y\}$ is known, the asymptotic dynamics of any initial state $\rho$ after sufficiently many steps $n$ is given by
\begin{equation}
\label{eq:asymptotic_behaviour}
\rho_{\infty} (n)\,= \sum\limits_{\lambda \in \sigma_{as},i} \lambda^n (X_{\lambda, i}, \rho)_\mathrm{HS}X_{\lambda, i}
\end{equation}
so that $\| \map R^n \rho - \rho_\infty (n) \| \to 0$ for $n \to \infty$.

Due to properties of the Haar measure of a group $\cal G$, a general twirling operation \eqref{eq:def_general_twirling} is an orthogonal projection. Therefore, the evolution generated by the RUO \eqref{eq:def_RUOs} converges to the desired twirling operation \eqref{eq:def_general_twirling} if and only if the asymptotic spectrum $\sigma_\mathrm{as}$ of $\map R$ contains solely the eigenvalue $\lambda =1$ and, simultaneously, the range of the twirling operation $\map T$ coincides with the set of fixed points of $\map R$, i.e.,
\begin{equation}
\label{eq:convergence_towards_twirling}
\set{Ran}\left(\map T\right) = \set{Attr}\left(\map R \right) = \set{Ker}\left(\map R-\mathds{1} \right).
\end{equation}
Moreover, the attractor eigenspaces $\set{Ker}\left(\map R - \lambda \mathds{1}\right)$ are determined by the set of attractor equations 
\begin{equation}
\label{eq:attractor_equations}
U_i X_{\lambda,i} = \lambda X_{\lambda,i}U_i,
\end{equation}
which must be satisfied by the attractors $X_{\lambda,i}$ simultaneously for all unitary operators $(U_i)_{i=1}^{m}$. The attractor equations \eqref{eq:attractor_equations} provide a convenient algebraic tool for deciding whether the convergence conditions \eqref{eq:convergence_towards_twirling} are fulfilled by a given set of unitary operators $\left(U_i\right)_{i=1}^m$ generating the RUO \eqref{eq:def_RUOs}. Note also that the assigned probabilities $(p_i)_{i=1}^{m}$ do not affect the resulting asymptotic dynamics of a RUO. This can be exploited to further speed up the convergence of the iterated dynamics towards the desired twirling operation. 
As a further consequence of the attractor equations \eqref{eq:attractor_equations} we obtain the result that the range of the twirling operation over the group ${\cal G} = \<U_1,...,U_m\>$ (i.e., the group generated by the unitary operators $U_1,...,U_m$) equals the set of fixed points of $\map R$, i.e., $\set{Ran}\left( \map T\right) = \set{Ker}\left(\map R-\mathds{1} \right)$. Therefore, the group $\cal G$ determines the asymptotic limit of the evolution generated by the RUO \eqref{eq:def_RUOs}, in the case that this limit exists.

Finally, we conclude that the iterative evolution \eqref{eq:def_iterative_approach} driven by  the RUO \eqref{eq:def_RUOs} approaches the twirling operation \eqref{eq:def_general_twirling} if and only if  all elements of the range of the twirling operation \eqref{eq:def_general_twirling} are solutions of the attractor equations \eqref{eq:attractor_equations} for $\lambda=1$ and there is no other solution. In the following  we apply this algebraic method to investigate which sets of unitaries are suitable to approximate the twirling of pairs of finite-dimensional qudits averaged over the group ${\cal U}(d) \otimes {\cal U}(d) \coloneqq \{U\otimes U\ |\ U \in {\cal U}(d) \} \subseteq {\cal U}(d^2)$.

\section{Twirling and Werner States} \label{sec:twirling_and_werner_states}
Entanglement distillation protocols \cite{BennettPurification1996} exploit twirling operations as one of the tools which brings a weakly entangled composite quantum system
into a Werner state by local operations and classical communication \cite{NielsenChuang}, i.e. 
without mutual interaction between the individual subsystems. 
In general, Werner states, introduced in \cite{Werner}, are mixed $N$-qudit states $\rho$, supported on the Hilbert space $\mathcal{H} \simeq (\mathbb{C}^d)^{\otimes N}$, which are left unaltered if all qudits involved undergo the same local unitary evolution, i.e.,
\begin{equation}
U^{\otimes N}\rho\  U^{\otimes N \dagger} = \rho
\end{equation}
for all unitary single-qudit operators $U \in {\cal U}(d)$. Correspondingly, these states can be obtained as twirled states resulting from the twirling operation
\begin{equation}
\label{eq:twirling_purification_qudit}
\map P (\rho) \coloneqq \int_{{\cal U}(d)} U^{\otimes N}\rho\ U^{\otimes N \dagger} \mathrm{d}U
\end{equation}
with the Haar measure $\mathrm{d}U$ on the unitary group ${\cal U}(d)$. Note that $\map P$ coincides with $\map T$ from Eq. \eqref{eq:def_general_twirling} for ${\cal G} = {\cal U}(d)^{\otimes N}$. 
Such a twirling operation maps any state $\rho$ to a Werner state and, conversely, every Werner state is contained in the range of $\map P$.
In the case of bipartite Hilbert spaces, the description of Werner states simplifies significantly \cite{Werner}. They form a one-parameter family of states which are a mixture of symmetric and antisymmetric states, i.e.,
\begin{equation}
\label{eq:werner_state}
\rho = \eta \frac{2}{d(d+1)} P_\mathrm{sym} + (1-\eta)\frac{2}{d(d-1)} P_\mathrm{asym}
\end{equation}
for $0 \leq \eta \leq 1$. Here $P_\mathrm{sym}$ and $P_\mathrm{asym}$ are the projections onto the symmetric and antisymmetric eigenspaces of the flip operator $F |\phi\>|\psi\> \coloneqq |\psi\>|\phi\>$ corresponding to the eigenvalues 1 and -1, respectively. 

\subsection{Twirling of Two Qubits}\label{sec:twirling_qubit}
The simplest situation arises for a twirling operation \eqref{eq:twirling_purification_qudit} acting on two qubits. 
In such a case we have $P_\mathrm{asym} = |\psi_-\>\<\psi_-|$ with the singlet state $\ket{\psi_-}\coloneqq 1/\sqrt{2}\left(|1\>|0\> - |0\>|1\>\right)$.
Consequently, according to condition \eqref{eq:convergence_towards_twirling}, asymptotically the two-qubit twirling operation \eqref{eq:twirling_purification_qudit} can be achieved by iteration of the RUO \eqref{eq:def_RUOs} if and only if
\begin{equation}
    \set{Attr}(\map R) = \set{span}\left\{\mathds{1},|\psi_{-}\>\<\psi_{-}| \right\}.
\end{equation}
Therefore, a natural question arises: what is the most general form of the set of unitary operators $\left(U_i\right)_{i=1}^m$ capable of asymptotically generating a Werner state? Employing the attractor equations \eqref{eq:attractor_equations}, we can provide an exhaustive answer. 

First of all let us focus on a scenario with only two unitary operators, say $U_1=u_1\otimes u_1$ and $U_2=u_2 \otimes u_2$. Without loss of generality $u_1, u_2 \in SU(2)$ and the computational basis of both qubit systems can be chosen in such a way that the matrix representation of one of these unitary single qubit operations, say $u_1$, is diagonal, i.e.,
\begin{eqnarray}\label{eq:KyselaParameters1}
u_1&=&\begin{bmatrix}
\mathrm{e}^{i \varphi} & 0 \\
0 & \mathrm{e}^{-i \varphi} \\
\end{bmatrix}
\end{eqnarray}
with $\varphi \in [0,2\pi)$, and that the matrix representation of the second unitary single qubit operation $u_2$ takes the general form
\begin{eqnarray}\label{eq:KyselaParameters2}
u_2&=&\begin{bmatrix}
\rm{e}^{i \theta} \rm{cos}(\gamma) & -\rm{e}^{-i\mu}\rm{sin}(\gamma) \\
\rm{e}^{i\mu}\rm{sin}(\gamma) & \rm{e}^{-i \theta}\rm{cos}(\gamma) \\
\end{bmatrix}
\end{eqnarray}
with $\theta, \mu\in [0,2\pi),~ \gamma \in [0,\pi)$.
By explicit solution of the attractor equations \eqref{eq:attractor_equations} it can be shown
that RUOs involving the corresponding unitary two-qubit operators $U_1$ and $U_2$
converge asymptotically to a Werner state for arbitrary probabilities $p_1, p_2 = (1-p_1)\in (0,1)$, if and only if the parameters
$\varphi, \gamma, \theta$
satisfy one of the following relations:
\begin{itemize}
	\item{$\varphi \in \{ \frac{\pi}{2},\frac{3\pi}{2}\}$ and $\gamma \notin \{ 0, \frac{\pi}{2}\}$ and $\theta \notin \{ 0, \frac{\pi}{2}, \pi, \frac{3\pi}{2}\}$;}
	\item{$\varphi \notin \{ \frac{\pi}{2},\frac{3\pi}{2}\}$ and $\gamma \notin \{ 0, \frac{\pi}{2}\}$;}
\end{itemize}
independently of $\mu$. For more than two unitary operators involved in the iterative generation of Werner states, i.e., $m > 2$, we can generalise the necessary and sufficient condition for convergence in the following way.
Again we can always choose a computational basis such that one
of our matrices, say $u_{i_0}$, is diagonal for some $i_0\in I \coloneqq \{1,...,m\}$
and is characterised by a single parameter, say $\varphi_{i_0}$, analogous to Eq. \eqref{eq:KyselaParameters1}.
Each of the remaining matrices $u_j$ with $i_0 \neq j \in I$ is parametrised
by three parameters $\gamma_j^{(i_0)}, \theta_j^{(i_0)}, \mu_j^{(i_0)}$ in analogy to equation \eqref{eq:KyselaParameters2}. 
 The iterative dynamics \eqref{eq:def_iterative_approach} generated by RUO \eqref{eq:def_RUOs} with $(U_i=u_i\otimes u_i)_{i=1}^{m}$ converge towards the two-qubit twirling operation \eqref{eq:twirling_purification_qudit} if and only if one of the following twirling conditions holds: 
\begin{itemize}
\item{If $\mathrm{Tr}(u_{i_0})\neq 0$ for some $i_0 \in I$, then there is some 
	$i_0\neq j \in I$, for which $\gamma_j^{(i_0)} \notin \{ 0, \frac{\pi}{2}\}$.}
\item{If $\mathrm{Tr}(u_{i})=0$ for all $i\in I$, we choose an arbitrary $i_0 \in I$ and associated $u_{i_0}$. 
Then there exist $j, k_1, k_2\in I \setminus \{i_0\}$ such that $\gamma_j^{(i_0)} \notin \{ 0, \frac{\pi}{2}\}$,
and $\mu_{k_1}^{(i_0)}\neq \mu_{k_2}^{(i_0)} +r \frac{\pi}{2}$ for all $r \in \mathbb{Z}$.}
\end{itemize}
The proof of both statements is straightforward but lengthy and is presented in Ref. \cite{Kysela-Thesis}. 
An immediate consequence of these statements is that for any random choice of operators $(u_i)_{i=1}^m$ the associated iterated RUO prepares Werner states asymptotically. 
Numerical evidence suggests that this is also the case for qudits with $d > 2$ (see Fig. \ref{fig:twirling_4d}).
Furthermore, the above mentioned characterisation of two-qubit twirling for RUOs involving an arbitrary number of unitary operators $(U_i)_{i=1}^m$ implies that any RUO \eqref{eq:def_RUOs} that prepares Werner states gives rise to another RUO that also prepares Werner states, using at most four of the original unitary operators $U_i$.

\begin{figure}[!htbp]
	\centering
	\includegraphics[width=0.5\textwidth]{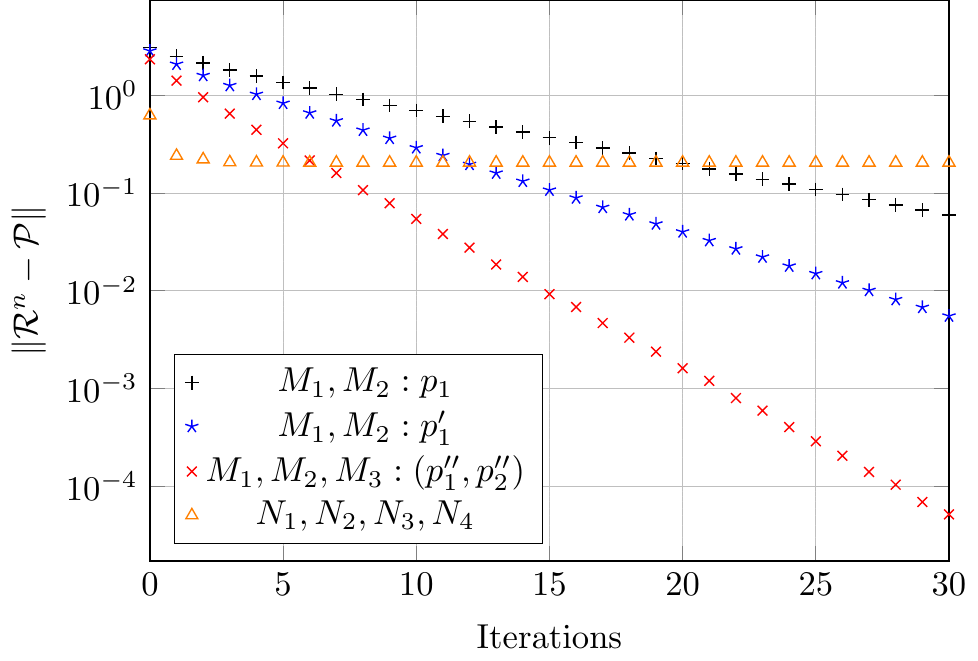}
	\caption{Hilbert-Schmidt distance between the two-qubit twirling operation \eqref{eq:twirling_purification_qudit} and the iterated RUO associated with the given unitary operators as a function of the number of iterations: The involved RUOs are given by two unitary operators $M_1,M_2$ with probabilities $p_1=0.75$ (black, plus); two unitary operators $M_1,M_2$ with optimised probabilities $p_1'= 0.459$ (blue, star); three unitary operators $M_1,M_2,M_3$ with optimised probabilities  $(p_1'',p_2'')=(0.41,0.18)$ (red, cross); and four unitary operators $(N_i)$ avoiding the twirling conditions (yellow, triangle). The Hilbert-Schmidt norm is plotted on a logarithmic scale. (For interpretation of the colours in the figure(s), the reader is referred to the web version of this article.)}
	\label{fig:twirling_qubits}
\end{figure}

In \cite{NovotnyRUOs} it was shown that the convergence of the iterated dynamics \eqref{eq:def_iterative_approach} generated by a RUO towards its asymptotic regime is exponential with the number of iterations. However, one can further improve the rate of convergence by properly chosen unitary operators $U_i$ and their associated probabilities $p_i$, provided they follow the twirling conditions. In Fig. \ref{fig:twirling_qubits} we compare exponential convergence rates of the iterated dynamics for different chosen settings with unitary operators $M_1, M_2$ and $M_3$ defined via parameters $\varphi_1=\frac{\pi}{4}, \theta_2=\frac{\pi}{4}, \mu_2=0, \gamma_2=\frac{\pi}{4}, \theta_3=0, \mu_3=\frac{\pi}{4}, \gamma_3=\frac{\pi}{4}$. This figure demonstrates how optimised probabilities or an extension of the set of unitary operators may significantly speed up the resulting convergence. Fig.  \ref{fig:twirling_qubits} also shows an example of a RUO whose four involved unitary operators $\left(N_i\right)$ do not follow the twirling conditions ($\mathrm{Tr}(N_1)\neq0$, and $\gamma_j^{(1)}\in\{0, \frac{\pi}{2}\}$ for all $j \in I \setminus \{1\}$).

\subsection{Twirling of Two Qudits}\label{sec:twirling_qudit}
It is a demanding task to derive
the analogous necessary and sufficient conditions which generalise the results of 
Sec. \ref{sec:twirling_qubit}
to two-qudit twirling operations \eqref{eq:twirling_purification_qudit}.
In this section we restrict ourselves to the less difficult but practically interesting problem of determining a generic set of three unitary operators capable of implementing the twirling operation \eqref{eq:twirling_purification_qudit} asymptotically for any finite dimension $d\geq 2$ of the underlying Hilbert space.
For this purpose 
let us consider an arbitrary finite-dimensional bipartite qudit system with Hilbert space $\mathcal{H} \simeq \mathbb{C}^d \otimes \mathbb{C}^d$ and let us fix some orthonormal basis $\ket{1},...,\ket{d}$ on both qudit subsystems. Furthermore, let $\map R$ be the RUO \eqref{eq:def_RUOs}
with unitary operators $U_i = u_i \otimes u_i$ for $u_1 = h, u_2 = u$ and $u_3 = v$, and let the relevant single-qudit unitary operators be defined by

\begin{itemize}
\item{$h \ket{k} \coloneqq \exp \left( 2^{k-d} \pi i \right) \ket{k}$},
\item{$u \ket{k} \coloneqq \ket{(k\ \mathrm{mod}\ d)+1}$},
\item{$v \coloneqq A \oplus \mathds{1}_{d-2}$, where $A \in \mathcal{U}(2)$ has no vanishing entries.}
\end{itemize}
We now prove that if $\map R$ is asymptotically stationary then this RUO prepares Werner states asymptotically, i.e., $\map R^n$ converges to $\map P$ \eqref{eq:twirling_purification_qudit}.

\begin{proof}
According to the equations \eqref{eq:convergence_towards_twirling} and \eqref{eq:werner_state}, it suffices to show that $\set{Ker}(\map R - \mathds{1}) = \set{span}\left\{P_\mathrm{sym},P_\mathrm{asym} \right\}$. From the definition of Werner states, it is clear that $\set{Ker}(\map R - \mathds{1}) \subseteq \set{span}\left\{P_\mathrm{sym},P_\mathrm{asym} \right\}$. In order to show the other inclusion, we introduce the orthonormal basis $B$ consisting of the vectors
\begin{align}\label{eq:twirling_basis}
\begin{split}
\ket{\phi_i} &\coloneqq \ket{i}\ket{i}, \\
\ket{\phi_{i,j}} &\coloneqq \frac{1}{\sqrt{2}}( \ket{i}\ket{j} + \ket{j}\ket{i} ), \\
\ket{\psi_{i,j}} &\coloneqq \frac{1}{\sqrt{2}}( \ket{i}\ket{j} - \ket{j}\ket{i} )
\end{split}
\end{align}
with $i<j\in \{1,...,d\}$. These states are the eigenbasis of the flip operator $F$ so that $P_\mathrm{sym}$ and $P_\mathrm{asym}$ are diagonal with respect to $B$. Thus, by equation \eqref{eq:werner_state}, it suffices to show that any eigenvector $X \in \set{Ker}(\map R - \mathds{1})$ is diagonal with respect to $B$ and that the diagonal matrix elements corresponding to $P_\mathrm{sym}$ and $P_\mathrm{asym}$ coincide, respectively. \\
If $X \in \set{Ker}(\map R - \mathds{1})$ the attractor equations \eqref{eq:attractor_equations}  imply the relations
\begin{equation*}
h^\dagger \otimes h^\dagger X\ h \otimes h = u^\dagger \otimes u^\dagger X\ u \otimes u = v^\dagger\otimes v^\dagger X\ v \otimes v = X.
\end{equation*}
As $h \otimes h$ is diagonal with respect to $B$, we can write $h \otimes h = \sum_{i=1}^d h_i \ket{b_i}\bra{b_i}$ with basis vectors $\ket{b_i} \in B$ and $|h_i| = 1$. We find
\begin{equation*}
\bra{b_i} X \ket{b_j} = \bra{b_i} h^\dagger\otimes h^\dagger X\ h \otimes h \ket{b_j} = h_i h_j^* \bra{b_i} X \ket{b_j}
\end{equation*}
for all $i$ and $j$, which implies that we have $\bra{b_i} X \ket{b_j} = 0$ whenever $h_i \neq h_j$. Using the choice of the diagonal entries of $h$, it follows that the only non-vanishing non-diagonal matrix elements of $X$ with respect to $B$ are $\bra{\phi_{i,j}} X \ket{\psi_{i,j}}$ and $\bra{\psi_{i,j}} X \ket{\phi_{i,j}}$ for $i<j$. \\
Moreover, since we have $u^\dagger\otimes u^\dagger X\ u \otimes u = X$, it follows that matrix elements of $X$ corresponding to the same orbit of $u \otimes u$, acting on the basis $B$, have to coincide. Note that each of these orbits contains at least one of the vectors in Eq. \eqref{eq:twirling_basis} with $i = 1$. Thus, it remains to show that $\bra{\phi_{1,k}}X\ket{\psi_{1,k}} = 0 = \bra{\psi_{1,k}}X\ket{\phi_{1,k}}$, $\bra{\phi_{1,k}}X\ket{\phi_{1,k}} = \bra{\phi_1}X\ket{\phi_1}$ and $\bra{\psi_{1,k}}X\ket{\psi_{1,k}} = \bra{\psi_{1,2}}X\ket{\psi_{1,2}}$ for all $1 < k \leq d$. Using a parametrisation of $A$ of the form
\begin{equation}\label{eq:u2_param}
A = \mathrm{e}^{i \varphi} \begin{bmatrix}
\alpha & \beta \\
-\overline{\beta} & \overline{\alpha} \\
\end{bmatrix},
\end{equation}
for $\alpha \neq 0 \neq \beta$ and $|\alpha|^2 + |\beta|^2 = 1$, the validity of these equations can be shown by straightforward calculation and induction.
\end{proof}

In general the RUO $\map R$ is not necessarily asymptotically stationary for all choices of $A \in {\cal U}(2)$. 
However, the 
attractor equations \eqref{eq:attractor_equations} imply that the condition $\bigcap_{i=1}^m \{ \lambda \lambda^\ast\ |\ \lambda \in \sigma(U_i) \} \subseteq \{1\}$ is sufficient for asymptotic stationarity of a RUO \eqref{eq:def_RUOs}, which can be used to show that
$\map R$ is asymptotically stationary for almost all choices of $A$, i.e., the set of parameters of $A$ for which $\map R$ is not guaranteed to be asymptotically stationary has measure zero \cite{Stonner}.
The presented construction depends on six continuous independent real-valued parameters, namely $p_1, p_2$ and the four parameters defining a particular unitary operator $A \in {\cal U}(2)$ (compare with Eq. \eqref{eq:u2_param}). 
Thus, this construction has the advantage that these parameters can be varied in order to optimise the convergence rate of the iterated RUO. Furthermore, since the eigenspace $\set{Ker}(\map R - \mathds{1})$ is completely determined by the (possibly countably infinite) group ${\cal G} = \braket{h, u, v}$ generated by the operators $h, u$ and $v$, this construction gives rise to a whole family of RUOs satisfying $\set{Ker}(\map R - \mathds{1}) = \set{span}\left\{P_\mathrm{sym},P_\mathrm{asym} \right\}$. 
It is also possible to generalise this construction without having to change the crucial arguments of the proof. For instance, the non-trivial part of $v$ could be defined on any other two-dimensional subspace of $\mathbb{C}^d$ or $u$ could be replaced by any other unitary operator that acts transitively on the fixed orthonormal basis, and one could introduce additional parameters into its definition.

\begin{figure}[!htbp]
  \centering
  \includegraphics[width=0.5\textwidth]{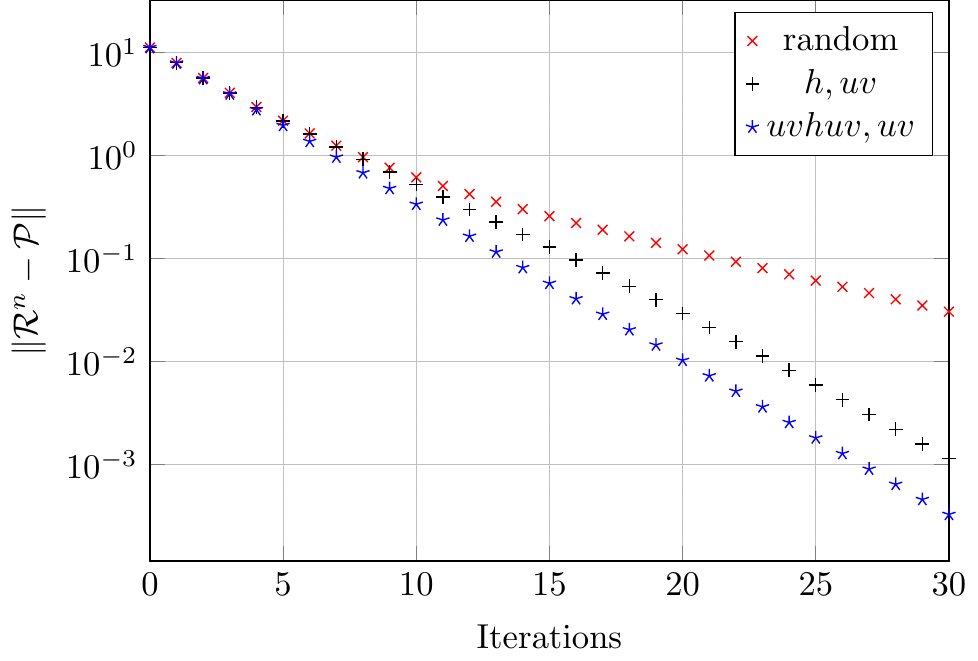}
  \caption{Hilbert-Schmidt distance between the two-qudit twirling operation \eqref{eq:twirling_purification_qudit} and the iterated RUO associated with the given unitary operators as a function of the number of iterations, in dimension $d = 4$: The involved RUOs are given by two randomly generated unitary operators (red, cross), the two unitary operators $h$ and $uv$ (black, plus) and the two unitary operators $uvhuv$ and $uv$ (blue, stars). The probability distribution $p_1, p_2 = 1-p_1$ has been optimised numerically in each case, in order to obtain the fastest possible convergence. The Hilbert-Schmidt norm is plotted on a logarithmic scale.}

  \label{fig:twirling_4d}
\end{figure}

If the dimension $d$ characterising the qudits is an odd number, then it is possible to show that the RUO $\map R_2$ involving only the two unitary operators $U_1 = h\otimes h$ and $U_2 = uv \otimes uv$ (according to Eq. \eqref{eq:def_RUOs}) also satisfies
$\ker(\map R_2 - \mathds{1}) = \set{span}\left\{P_\mathrm{sym},P_\mathrm{asym} \right\}$. Thus, if it is asymptotically stationary then $\map R_2$ prepares Werner states asymptotically as well \cite{Stonner}. Even though all the crucial arguments of the above proof stay valid in such a case, the calculations get much more involved. Numerical evidence suggests the conjecture that this construction may also work in even dimensions. 
Indeed, Fig. \ref{fig:twirling_4d} shows the convergence of the iterated dynamics (\ref{eq:def_iterative_approach}) generated by the RUO $\map R_2$ in dimension $d = 4$. 
It is compared with the iterated dynamics generated by RUO \eqref{eq:def_RUOs} with two randomly chosen unitary operators $u_1, u_2 \in {\cal U}(4)$ and $U_i = u_i \otimes u_i$, as well as with a RUO with $u_1 = uvhuv$ and $u_2 = uv$. Note that the generated groups $\braket{uvhuv,uv}$ and $\braket{h,uv}$ coincide.
In order to obtain fast convergence, the values of the probabilities $p_i$ and the parameters $\alpha, \beta, \varphi$ according to Eq. \eqref{eq:u2_param} have been optimised using a sequential least squares algorithm for a large number of randomly chosen initial values. The figure demonstrates the significant advantage, in terms of convergence rates, of optimising continuous parameters instead of choosing them at random.

\section{Conclusions}\label{sec:conclusion}
Werner states are of considerable interest in the field of quantum information and have useful practical applications in the important area of entanglement purification \cite{BennettPurification1996,Delgado}. It is therefore of great utility to be able to generate these states in a manner which is both experimentally easy to implement and quick to converge.

In this work we give a general recipe for how to choose unitary operators which, applied randomly and repeatedly, efficiently approximate a given twirling operation. This method is applied to investigate procedures allowing us to generate Werner states using random unitary operations. 
Firstly, we found a set of criteria which any set of unitary operators $\left(U_i\right)_{i=1}^{m}$ generating a RUO must satisfy in order to bring a two-qubit system into a Werner state asymptotically for arbitrary $m\geq 2$. Secondly, we formulated a procedure for generating bipartite Werner states in arbitrary finite dimension $d\geq 2$ using RUOs. 
This can be achieved using only three unitary operators, and in fact there are infinitely many such triplets which yield this result. 
Furthermore, if $d$ is odd we only need two unitary operators. There is numerical evidence to conjecture that this is also true for even dimensions $d$.

It is shown that various RUOs may be chosen to generate the same desired twirling operation exponentially quickly with the number of iterations. By adjusting parameters specifying unitary operators and their associated probabilities one can significantly improve the convergence rate.
A systematic exploration of the issue of convergence rates is beyond the scope of this paper, although some first results are presented here. Another avenue which might be explored in the future on the basis of our results is the question of whether RUOs might be generalised to generate other states, for instance isotropic states - i.e., bipartite qudit states which are invariant under the operation $U\otimes U^{\ast} (.) ~U^{\dagger}\otimes U^{\ast\dagger}$ for all $U \in {\cal U}(d)$ acting on one qudit \cite{AlberBook, Terhal2000, Rungta2003}.

{\it Acknowledgements} JN, GW and IJ acknowledge the financial support from M\v{S}MT No. 8J18DE011, RVO14000 and ``Centre for Advanced Applied Sciences'', Registry No. CZ.02.1.01/0.0/0.0/16\_019/0000778, supported by the Operational Programme Research, Development and Education, co-financed by the European Structural and Investment Funds and the state budget of the Czech Republic. JN and IJ are supported by the Czech Science foundation (GACR) project number 16-09824S. IJ is partially supported from GACR 17-00844S.
JS, JN and GA acknowledge support by the DAAD (PPP-Tschechien, 57390900) and by the Center of Excellence ``Crossing'' funded by the DFG.

\end{document}